\documentclass[conference]{IEEEtran}

\usepackage{amsmath}
\usepackage{graphicx}
\usepackage{color}
\usepackage{cite}

\begin{document}
	
	
	\graphicspath{{images/}}

	\title{3-D Motion Correction for Volumetric Super-Resolution Ultrasound (SR-US) Imaging}
	
	\author{\IEEEauthorblockN{Sevan~Harput$^1$, Kirsten Christensen-Jeffries$^2$, Jemma Brown$^2$, Jiaqi Zhu$^1$, Ge Zhang$^1$, \\ Robert J. Eckersley$^{2*}$, Chris Dunsby$^{3*}$, and Meng-Xing~Tang$^{1*}$}
		\IEEEauthorblockA{ \small \\
			\small $^1$ULIS Group, Department of Bioengineering, Imperial College London, London, SW7 2BP, UK  \\
			\small $^2$Biomedical Engineering Department, Division of Imaging Sciences, King's College London, SE1 7EH, London, UK  \\
			\small $^3$Department of Physics and the Centre for Pathology, Imperial College London, London, SW7 2AZ, UK  \\
			\small $^*$These authors contributed equally to this work  \\
			E-mail: S.Harput@imperial.ac.uk, Mengxing.Tang@imperial.ac.uk}	
	}
	
	\maketitle

	\begin{abstract}
		\boldmath
Motion during image acquisition can cause image degradation in all medical imaging modalities. This is particularly relevant in 2-D ultrasound imaging, since out-of-plane motion can only be compensated for movements smaller than elevational beamwidth of the transducer. Localization based super-resolution imaging creates even a more challenging motion correction task due to the requirement of a high number of acquisitions to form a single super-resolved frame.

In this study, an extension of two-stage motion correction method is proposed for 3-D motion correction. Motion estimation was performed on high volumetric rate ultrasound acquisitions with a handheld probe. The capability of the proposed method was demonstrated with a 3-D microvascular flow simulation to compensate for handheld probe motion. Results showed that two-stage motion correction method reduced the average localization error from 136 to 18~$\mu$m.
				
	\end{abstract}
	
	\maketitle

	\section{Introduction}
	
Localization based super-resolution ultrasound (SR-US) imaging can overcome the diffraction resolution limit~\cite{Couture2011,Viessmann2013,Desailly2013}. However, SR-US has two key restrictions on the maximum achievable resolution. Firstly, the SNR of the imaging system must be sufficiently high and the microbubble localization method must be able to identify individual microbubbles. The second limitation is any error introduced by motion during image acquisition~\cite{Harput2017a,Harput2018}. SR-US imaging and other imaging modalities based on multiple acquisitions are prone to motion artefacts that can usually be compensated by applying motion correction algorithms. For 2-D imaging, motion correction is impossible in the elevational direction in the presence of out-of-plane motion. For 3-D imaging with plane waves, the motion can be compensated in every direction thanks to the availability of full volumetric information at each time-point in the high-speed acquisition.

 SR-US has been demonstrated by several researchers using different imaging methods and experimental setups. The most common way of achieving SR-US images is by using a 1-D ultrasound probe and super-localizing microbubbles in 2-D~\cite{Christensen-Jeffries2015,Ackermann2016,Foiret2017,Harput2017b,Harput2018,Couture2018,Song2018,Opacic2018,Ilovitsh2018,Zhang2018}. There are several studies that extended the use of SR-US to the third dimension by mechanically scanning a volume with a linear probe~\cite{Errico2015,Lin2017}. To achieve super-resolution in the elevational direction, researchers performed 3-D super-localization by using different methods~\cite{Reilly2013,Desailly2013,Christensen-Jeffries2017a}. So far, 3-D SR-US imaging has not been achieved in an \textit{in vitro}, pre-clinical or clinical setup involving large scale motion.

 The aim of this study is to reduce the error introduced by motion in SR-US imaging by using a 3-D extension of the two-stage motion correction method~\cite{Harput2017a,Harput2018}. To acquire realistic motion, a 2-D handheld probe was used to image a wire phantom with a volumetric imaging rate of 333 Hz. Three set of simulations were performed without motion, with motion, and with motion correction. The extracted motion was implemented in 3-D simulations of microbubbles flowing inside a microvessel through tissue. 3-D SR-US imaging was achieved in simulations in the presence of handheld probe motion.

	\section{Materials and Methods}

\subsection{3-D Motion Estimation \& Correction}

The motion estimation used here is based on an image registration approach which was previously proposed for \textit{in vivo} 2-D SR-US imaging~\cite{Harput2018}. MATLAB (The MathWorks, Natick, MA) codes are currently available to download~\cite{MC_url}. This method is capable of performing {rigid}, {affine}, {non-rigid}, and {two-stage} motion estimations. In this study, {affine} and {two-stage} methods achieved the same results due to the characteristics of handheld probe motion that was detected from the wire phantom acquisitions and simulated in the flow phantom.

Motion estimation was performed as the first step of the processing chain on the volumetric B-mode images. The transformation matrix was used to correct the 3-D microbubble localizations.

\subsection{Handheld Probe Measurements}

High volume-rate 3-D ultrasound imaging was performed with two ULA-OP256 systems~\cite{Boni2016,Boni2017}, which were used to transmit and receive synchronously from a 2-D sparse array. The 2-D sparse array was designed with 512 elements according to the method described in ~\cite{Ramalli2015a,Harput2018a}. The sparse array was used to image a 100~$\mu$m wire phantom using 3 cycle Gaussian pulses with a center frequency of 3.7~MHz. A total of 333 volumes were recorded using 9-angle plane wave compounding within a range of $\pm10$ degrees in the lateral and elevational directions with a pulse repetition frequency of 3000 Hz. During the 1 second acquisition, the probe was handheld as steadily as possible.

\subsection{3-D microvascular flow simulation}

\begin{figure}[!t]
	\centering
	\includegraphics[viewport = 30 55 450 650,  width = 54mm, clip]{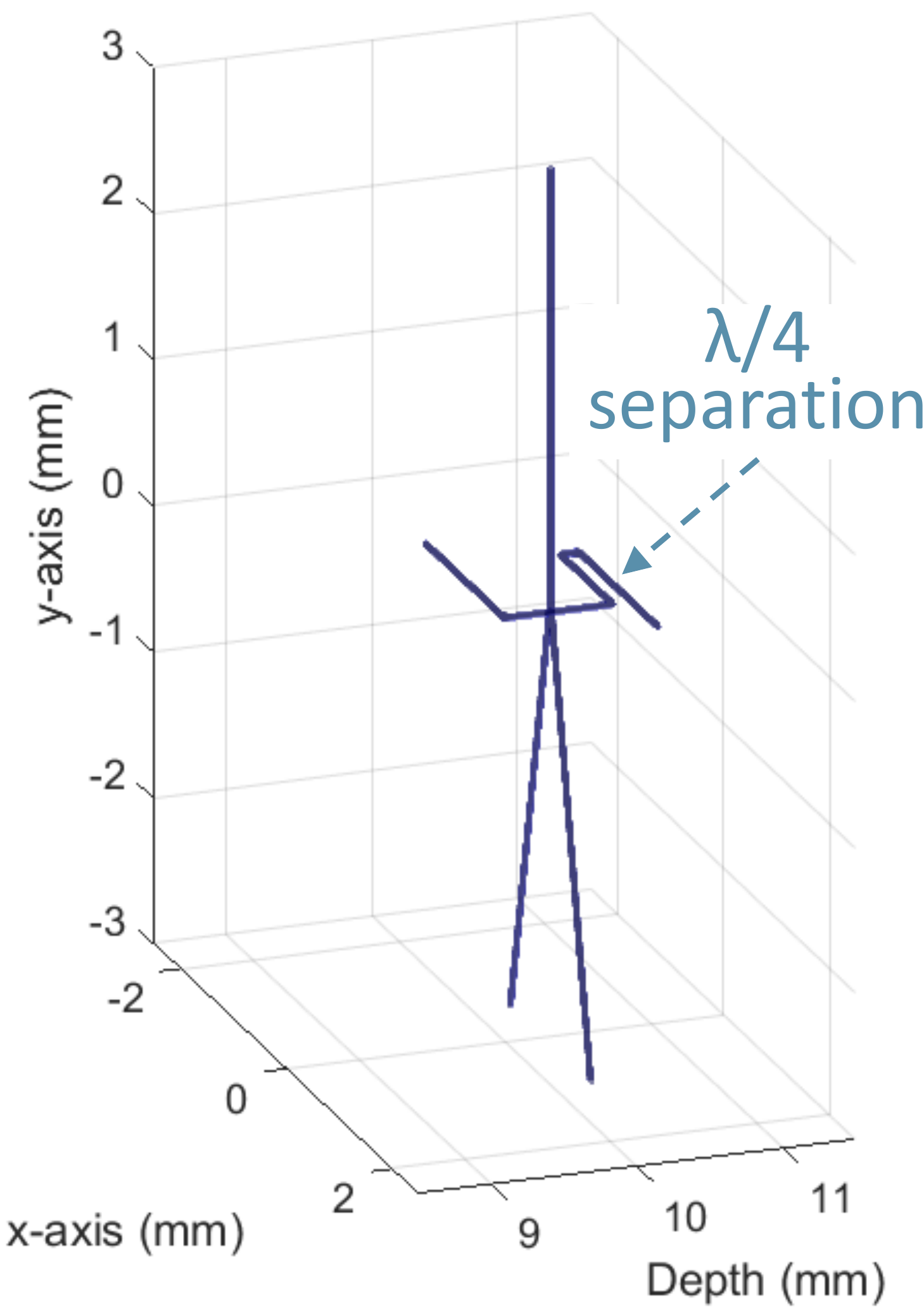}
	\caption{Shape of the simulated microvascular structure is shown in the figure. It is a combination of Y and S shaped microvessels with sub-wavelength features.}
	\label{fig:Simulation_microvessel}
\end{figure}

A 3-D microvascular flow simulation environment was created by combining tissue echoes generated in Field II~\cite{Jensen1992,Jensen1996} and microbubble signals generated with the Marmottant model~\cite{Marmottant2005}. Simulated microbubbles with a diameter of 3~$\mu$m  were placed at different locations inside a Y and S shaped microvessel phantom as shown in Fig.~\ref{fig:Simulation_microvessel}. To test the feasibility of motion correction for 3-D SR-US imaging, the microvessels were simulated with sub-wavelength structures, where the wavelength is 416~$\mu$m at the imaging frequency in average human tissue. Fig.~\ref{fig:Simulation_microvessel} only shows the simulated Y and S shaped microvessels for clear visualization. The rest of the simulated volume had scatterers to generate a fully developed speckle pattern. By changing the location of microbubbles in both tubes a constant stream was created within a velocity of 3 and 10~mm/s for the S and Y shaped vessels respectively. A total of 333 volumetric ultrasound frames were simulated. Motion extracted from the handheld probe measurements were used to move the location of scatterers and microbubbles spatially in every simulated volume. Singular value decomposition was used on the simulated data to separate the tissue and microbubble signals. Motion estimation was performed on the 3-D B-mode data and microbubble locations were corrected accordingly. Localization of isolated microbubbles was performed using the \textit{onset} method on every acquired volume to generate the 3-D SR-US images~\cite{Christensen-Jeffries2017}.

	\section{Results \& Discussion}

\begin{figure}[!t]
	\centering
	\includegraphics[viewport = 20 4 630 342,  width = 88mm, clip]{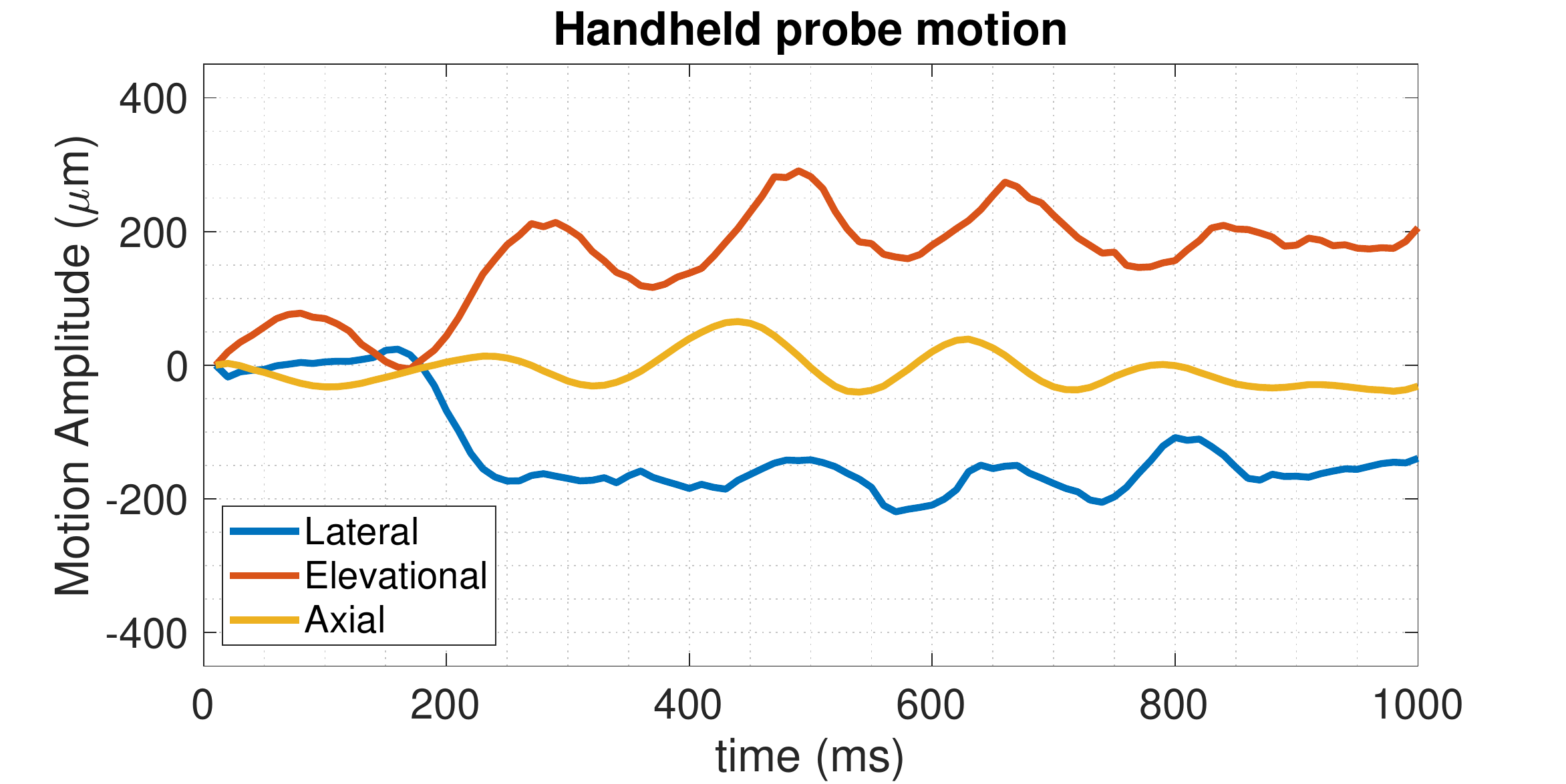}
	\caption{Handheld probe motion is estimated in three dimensions over a 1 second acquisition with 333 volumes.}
	\label{fig:Handheld_probe_motion}
\end{figure}

Handheld probe motion was estimated using the two-stage motion estimation method. The absolute motion in the axial, lateral, and elevational directions over 1 second is plotted as a function of time in Fig.~\ref{fig:Handheld_probe_motion}. The absolute motion was 384~$\mu$m, which is smaller than the imaging wavelength but much larger than typical localization precisions achieved in SR-US imaging.

\begin{figure*}[!t]
	\centering
	\includegraphics[viewport = 80 50 1130 550,  width = 175mm, clip]{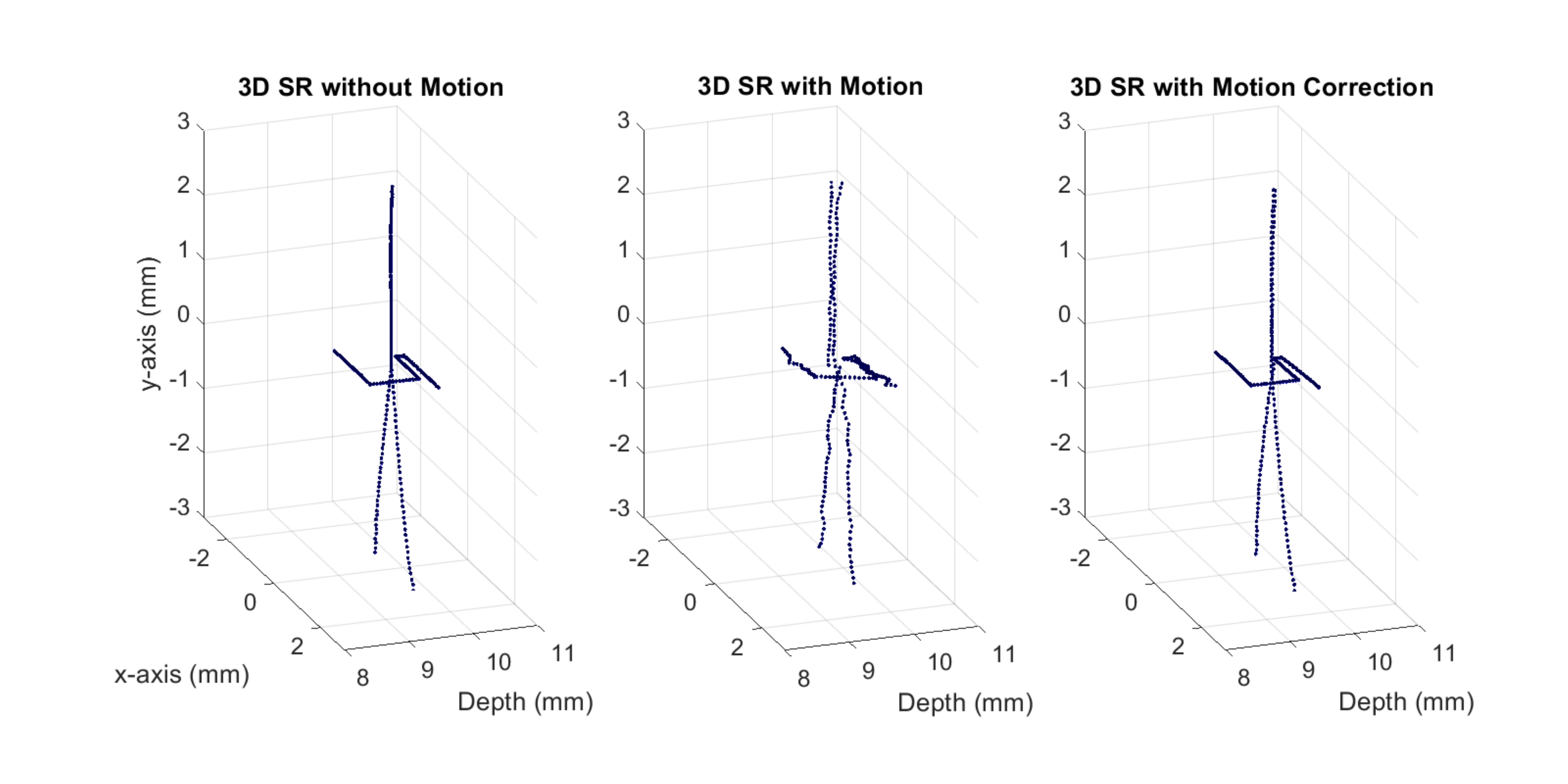}
	\caption{Figure shows the 3-D SR-US images achieved by super-localizing microbubbles flowing through the Y and S shaped microvessels (Left) without motion, (Middle) with motion, and (Right) with motion correction.}
	\label{fig:SR_comparison_MC}
\end{figure*}

Fig.~\ref{fig:SR_comparison_MC} shows the 3-D SR-US images achieved by super-localizing microbubbles flowing through the Y and S shaped microvessels in three different simulations without motion, with motion, and with motion correction, respectively. Without motion, the microvascular structures can be imaged clearly by using 3-D SR-US as shown in Fig.~\ref{fig:SR_comparison_MC} (left). In Fig.~\ref{fig:SR_comparison_MC} (middle), the S shaped vessel cannot be resolved due to motion artefacts and the main branch of the Y shaped vessel appears as two different vessels. The reason for this duplicated vessel is the two different microbubbles flowing through the main branch of the Y shape vessel at different times and therefore with different positions between the simulated phantom and probe. After correcting the positions of localized microbubbles, both Y and S shaped vessels were visualized without major motion artefacts in Fig.~\ref{fig:SR_comparison_MC} (right).

\begin{figure}[!t]
	\centering
	\includegraphics[viewport = 60 25 380 690,  width = 56mm, clip]{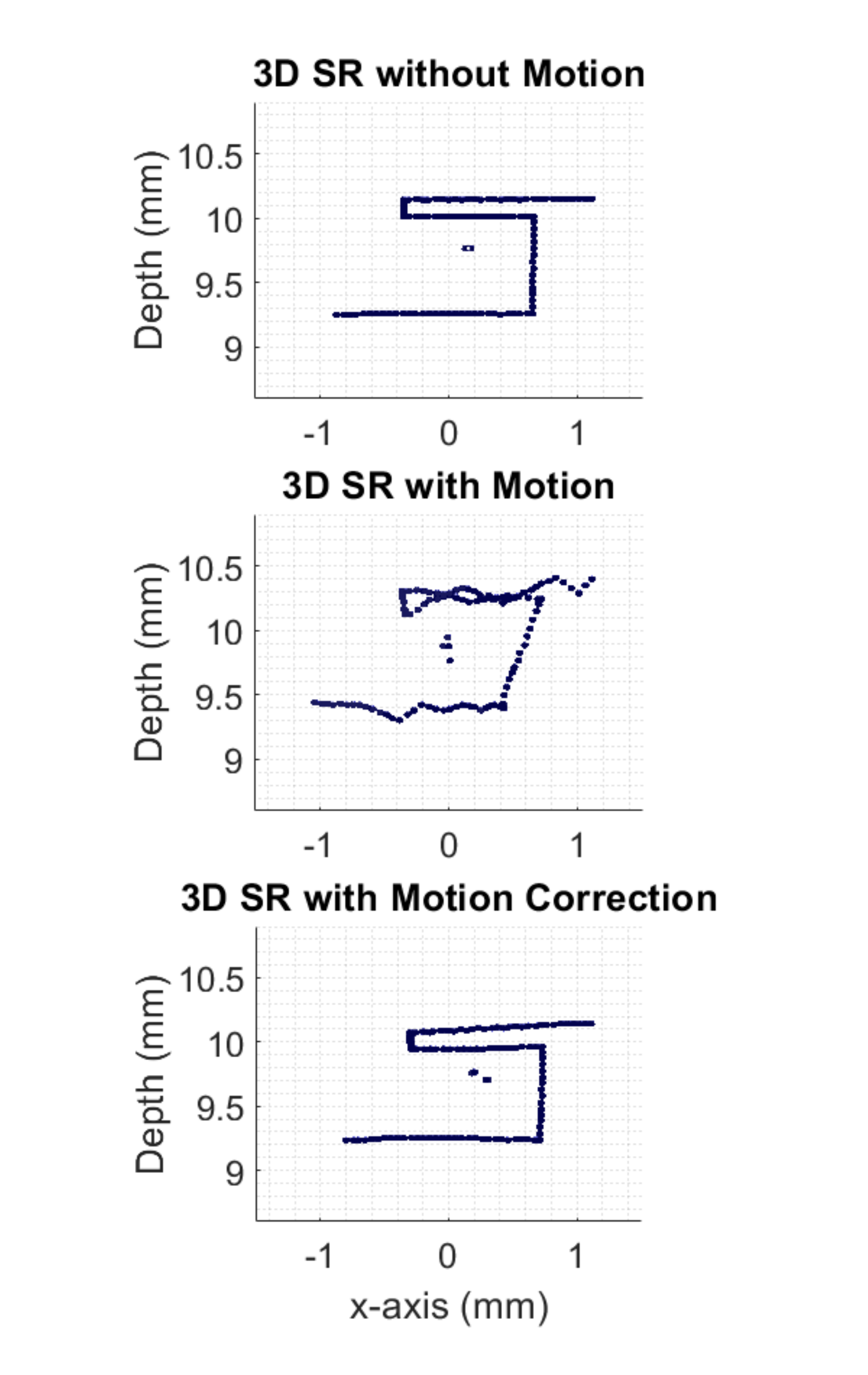}
	\caption{Figure plots a detailed view of the S shaped vessel to highlight the motion artefacts before and after motion correction.}
	\label{fig:SR_comparison_MC_S_vessel}
\end{figure}

\begin{figure}[!t]
	\centering
	\includegraphics[viewport = 1 30 420 490,  width = 60mm, clip]{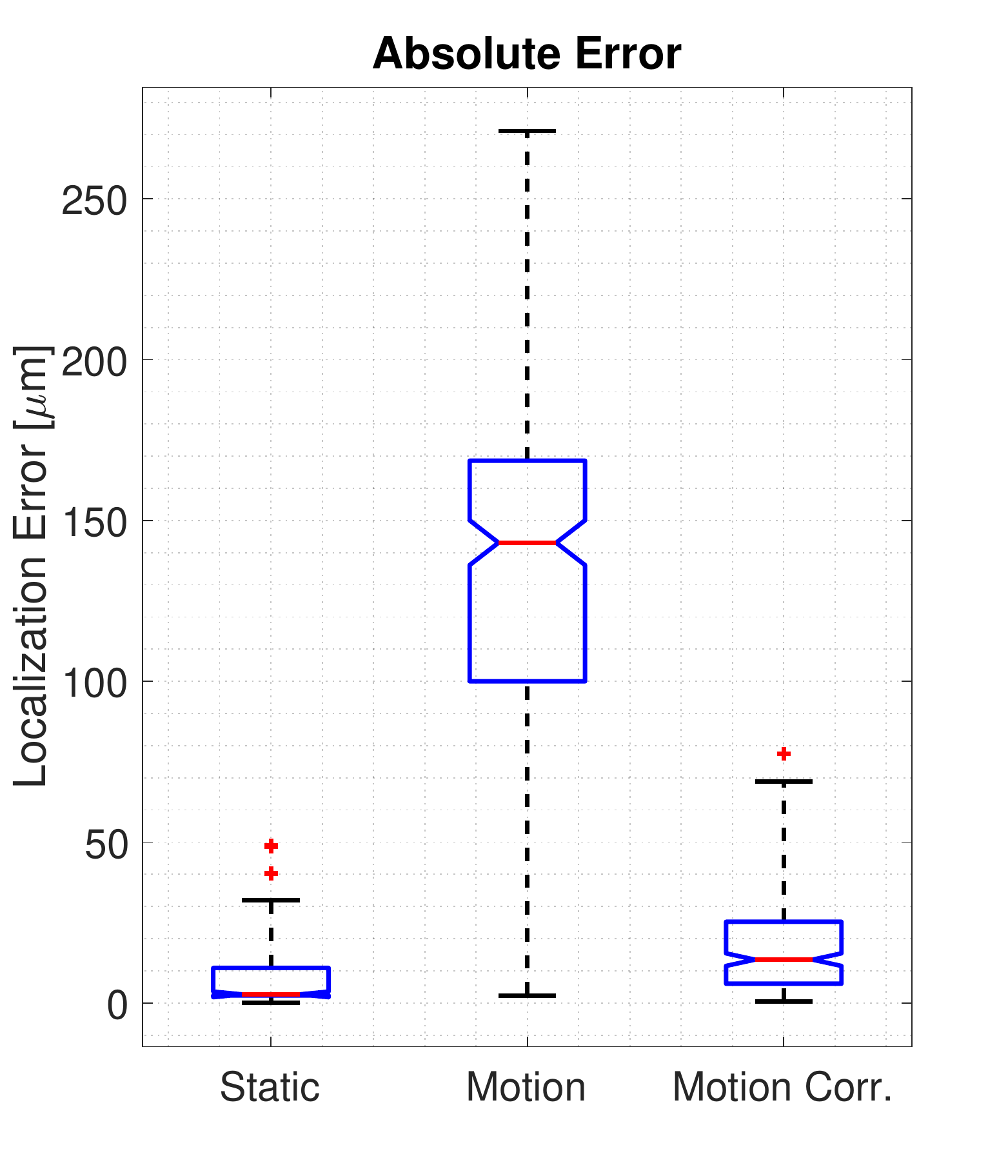}
	\caption{Localization error of microbubbles inside the S shaped vessel without motion, with motion, and after applying motion correction.}
	\label{fig:MC_error_Boxplot}
\end{figure}

Although, the main structure of both vessels was recovered after motion correction, residual motion artefacts can be seen when zoomed in to the S shaped vessel with sub-wavelength features in Fig.~\ref{fig:SR_comparison_MC_S_vessel}. The error values for the localizations of microbubbles travelling through the S shaped microvessel are presented as a box chart in Fig.~\ref{fig:MC_error_Boxplot} for three sets of simulations. Average absolute error between the simulated microbubble locations and 3-D super-localizations was 7.5, 136, and 18~$\mu$m for simulations without motion, with motion, and with motion correction. Motion correction significantly improved the erroneous localizations, where 75\% of the microbubble locations were detected with an error smaller than 25~$\mu$m. This value was 169~$\mu$m for the simulations with motion and 11~$\mu$m for the simulations without motion. Motion correction also reduced the maximum error below 78~$\mu$m, which is smaller than $\lambda /5$. A significant reduction in motion artefacts in the motion compensated 3-D SR-US clearly presents the benefit of motion correction.

	\section{Conclusion}

Handheld probe motion over a 1 second duration may not be large enough to make a visible difference in the B-mode images; however, it is large enough to generate motion artefacts in 3-D SR-US as shown in simulations. In a clinical setup, the motion artefacts might be more severe since tissue will generate more complicated motion patterns than a handheld probe alone. Motion is a problem for both 2-D and 3-D SR-US and reduces image quality and resolution, therefore there is a need for motion correction in SR-US imaging.

Unlike 2-D imaging, motion can be compensated in every direction thanks to the availability of full 3-D volumetric information. In this study, a 3-D extension of the two-stage motion correction algorithm was used to compensate for the probe motion. The capability of the motion correction algorithm was demonstrated via 3-D simulations of sub-wavelength structures.

	\section*{Acknowledgments}  
	   
This work was supported mainly by the EPSRC under Grant EP/N015487/1 and EP/N014855/1, in part by the King's College London (KCL) and Imperial College London EPSRC Centre for Doctoral Training in Medical Imaging (EP/L015226/1), in part by the Wellcome EPSRC Centre for Medical Engineering at KCL (WT 203148/Z/16/Z), in part by the Department of Health through the National Institute for Health Research comprehensive Biomedical Research Center Award to Guy's and St Thomas' NHS Foundation Trust in partnership with KCL and King's College Hospital NHS Foundation Trust, in part by the Graham-Dixon Foundation and in part by NVIDIA GPU grant.

	
\bibliography{BuBBle,Ultrasound,SignalProcessing,MotionCorrection,SuperRes,3D_Imaging}       

\begin{thebibliography}{10}
\providecommand{\url}[1]{#1}
\csname url@samestyle\endcsname
\providecommand{\newblock}{\relax}
\providecommand{\bibinfo}[2]{#2}
\providecommand{\BIBentrySTDinterwordspacing}{\spaceskip=0pt\relax}
\providecommand{\BIBentryALTinterwordstretchfactor}{4}
\providecommand{\BIBentryALTinterwordspacing}{\spaceskip=\fontdimen2\font plus
\BIBentryALTinterwordstretchfactor\fontdimen3\font minus
  \fontdimen4\font\relax}
\providecommand{\BIBforeignlanguage}[2]{{%
\expandafter\ifx\csname l@#1\endcsname\relax
\typeout{** WARNING: IEEEtran.bst: No hyphenation pattern has been}%
\typeout{** loaded for the language `#1'. Using the pattern for}%
\typeout{** the default language instead.}%
\else
\language=\csname l@#1\endcsname
\fi
#2}}
\providecommand{\BIBdecl}{\relax}
\BIBdecl

\bibitem{Couture2011}
O.~Couture, B.~Besson, G.~Montaldo, M.~Fink, and M.~Tanter, ``Microbubble
  ultrasound super-localization imaging (musli),'' in \emph{IEEE International
  Ultrasonics Symposium (IUS)}, 2011, pp. 1285--1287.

\bibitem{Viessmann2013}
O.~M. Viessmann, R.~J. Eckersley, K.~Christensen-Jeffries, M.-X. Tang, and
  C.~Dunsby, ``Acoustic super-resolution with ultrasound and microbubbles,''
  \emph{Phys. Med. Biol}, vol.~58, pp. 6447--6458, 2013.

\bibitem{Desailly2013}
Y.~Desailly, O.~Couture, M.~Fink, and M.~Tanter, ``Sono-activated ultrasound
  localization microscopy,'' \emph{Applied Physics Letters}, vol. 103, no.
  174107, 2013.

\bibitem{Harput2017a}
S.~Harput, K.~Christensen-Jeffries, Y.~Li, J.~Brown, R.~J. Eckersley,
  C.~Dunsby, and M.-X. Tang, ``Two stage sub-wavelength motion correction in
  human microvasculature for ceus imaging,'' in \emph{IEEE International
  Ultrasonics Symposium (IUS)}, 2017, pp. 1--4.

\bibitem{Harput2018}
S.~Harput, K.~Christensen-Jeffries, J.~Brown, Y.~Li, K.~J. Williams, A.~H.
  Davies, R.~J. Eckersley, C.~Dunsby, and M.~Tang, ``Two-stage motion
  correction for super-resolution ultrasound imaging in human lower limb,''
  \emph{IEEE Trans. Ultrason., Ferroelectr., Freq. Control}, vol.~65, no.~5,
  pp. 803--814, 2018.

\bibitem{Christensen-Jeffries2015}
K.~Christensen-Jeffries, R.~J. Browning, M.-X. Tang, C.~Dunsby, and R.~J.
  Eckersley, ``In vivo acoustic super-resolution and super-resolved velocity
  mapping using microbubbles,'' \emph{IEEE Trans. Med. Imag.}, vol.~34, no.~2,
  pp. 433--440, 2015.

\bibitem{Ackermann2016}
D.~Ackermann and G.~Schmitz, ``Detection and tracking of multiple microbubbles
  in ultrasound b-mode images,'' \emph{IEEE Trans. Ultrason., Ferroelectr.,
  Freq. Control}, vol.~63, no.~1, pp. 72--82, 2016.

\bibitem{Foiret2017}
J.~Foiret, H.~Zhang, T.~Ilovitsh, L.~Mahakian, S.~Tam, and K.~W. Ferrara,
  ``Ultrasound localization microscopy to image and assess microvasculature in
  a rat kidney,'' \emph{Scientific Reports}, vol.~7, no. 13662, pp. 1--12,
  2017.

\bibitem{Harput2017b}
S.~Harput, K.~Christensen-Jeffries, J.~Brown, R.~J. Eckersley, C.~Dunsby, and
  M.-X. Tang, ``Localisation of multiple non-isolated microbubbles with
  frequency decomposition in super-resolution imaging,'' in \emph{IEEE
  International Ultrasonics Symposium (IUS)}, 2017, pp. 1--4.

\bibitem{Couture2018}
O.~Couture, V.~Hingot, B.~Heiles, P.~Muleki-Seya, and M.~Tanter, ``Ultrasound
  localization microscopy and super-resolution: A state of the art,''
  \emph{IEEE Trans. Ultrason., Ferroelectr., Freq. Control}, vol.~65, no.~8,
  pp. 1304--1320, 2018.

\bibitem{Song2018}
P.~Song, J.~D. Trzasko, A.~Manduca, R.~Huang, R.~Kadirvel, D.~F. Kallmes, and
  S.~Chen, ``Improved super-resolution ultrasound microvessel imaging with
  spatiotemporal nonlocal means filtering and bipartite graph-based microbubble
  tracking,'' \emph{IEEE Trans. Ultrason., Ferroelectr., Freq. Control},
  vol.~65, no.~2, pp. 149--167, 2018.

\bibitem{Opacic2018}
T.~Opacic, S.~Dencks, B.~Theek, M.~Piepenbrock, D.~Ackermann, A.~Rix,
  T.~Lammers, E.~Stickeler, S.~Delorme, G.~Schmitz, and F.~Kiessling, ``Motion
  model ultrasound localization microscopy for preclinical and clinical
  multiparametric tumor characterization,'' \emph{Nature Communications},
  vol.~9, no. 1527, pp. 1--13, 2018.

\bibitem{Ilovitsh2018}
T.~Ilovitsh, A.~Ilovitsh, J.~Foiret, B.~Z. Fite, and K.~W. Ferrara,
  ``Acoustical structured illumination for super-resolution ultrasound
  imaging,'' \emph{Communications Biology}, vol.~1, no.~3, 2018.

\bibitem{Zhang2018}
G.~Zhang, S.~Harput, S.~Lin, K.~Christensen-Jeffries, C.~H. Leow, J.~Brown,
  C.~Dunsby, R.~J. Eckersley, and M.-X. Tang, ``Acoustic wave sparsely
  activated localization microscopy (awsalm): Super-resolution ultrasound
  imaging using acoustic activation and deactivation of nanodroplets,''
  \emph{Applied Physics Letters}, vol. 113, no.~1, p. 014101, 2018.

\bibitem{Errico2015}
C.~Errico, J.~Pierre, S.~Pezet, Y.~Desailly, Z.~Lenkei, O.~Couture, and
  M.~Tanter, ``Ultrafast ultrasound localization microscopy for deep
  super-resolution vascular imaging,'' \emph{Nature}, vol. 527, pp. 499--507,
  2015.

\bibitem{Lin2017}
F.~Lin, S.~E. Shelton, D.~Espindola, J.~D. Rojas, G.~Pinton, and P.~A. Dayton,
  ``3-d ultrasound localization microscopy for identifying microvascular
  morphology features of tumor angiogenesis at a resolution beyond the
  diffraction limit of conventional ultrasound,'' \emph{Theranostics}, vol.~7,
  no.~1, pp. 196--204, 2017.

\bibitem{Reilly2013}
M.~A. O'Reilly and K.~Hynynen, ``A super-resolution ultrasound method for brain
  vascular mapping,'' \emph{Medical Physics}, vol.~40, no. 110701, 2013.

\bibitem{Christensen-Jeffries2017a}
K.~Christensen-Jeffries, J.~Brown, P.~Aljabar, M.-X. Tang, C.~Dunsby, and R.~J.
  Eckersley, ``3-d in vitro acoustic super-resolution and super-resolved
  velocity mapping using microbubbles,'' \emph{IEEE Trans. Ultrason.,
  Ferroelectr., Freq. Control}, vol.~64, no.~10, pp. 1478--1486, 2017.

\bibitem{MC_url}
\BIBentryALTinterwordspacing
D.-J. Kroon. [Online]. Available: \url{https://uk.mathworks.com/
  matlabcentral/fileexchange/20057-b-spline-grid--image-and-point-based-registration}
\BIBentrySTDinterwordspacing

\bibitem{Boni2016}
E.~Boni, L.~Bassi, A.~Dallai, F.~Guidi, V.~Meacci, A.~Ramalli, S.~Ricci, and
  P.~Tortoli, ``Ula-op 256: A 256-channel open scanner for development and
  real-time implementation of new ultrasound methods,'' \emph{IEEE Trans.
  Ultrason., Ferroelectr., Freq. Control}, vol.~63, no.~10, pp. 1488--1495,
  2016.

\bibitem{Boni2017}
E.~Boni, L.~Bassi, A.~Dallai, V.~Meacci, A.~Ramalli, M.~Scaringella, F.~Guidi,
  S.~Ricci, and P.~Tortoli, ``Architecture of an ultrasound system for
  continuous real-time high frame rate imaging,'' \emph{IEEE Trans. Ultrason.,
  Ferroelectr., Freq. Control}, vol.~64, no.~9, pp. 1276--1284, 2017.

\bibitem{Ramalli2015a}
A.~Ramalli, E.~Boni, A.~S. Savoia, and P.~Tortoli, ``Density-tapered spiral
  arrays for ultrasound 3-d imaging,'' \emph{IEEE Trans. Ultrason.,
  Ferroelectr., Freq. Control}, vol.~62, no.~8, pp. 1580--1588, 2015.

\bibitem{Harput2018a}
S.~Harput, K.~Christensen-Jeffries, J.~Brown, J.~Zhu, G.~Zhang, C.~H. Leow,
  M.~Toulemonde, A.~Ramalli, E.~Boni, P.~Tortoli, R.~J. Eckersley, C.~Dunsby,
  and M.-X. Tang, ``3-d super-resolution ultrasound imaging using a 2-d sparse
  array with high volumetric imaging rate,'' in \emph{IEEE International
  Ultrasonics Symposium (IUS)}, 2018, pp. 1--4.

\bibitem{Jensen1992}
J.~Jensen and N.~B. Svendsen, ``Calculation of pressure fields from arbitrarily
  shaped, apodized, and excited ultrasound transducers,'' \emph{IEEE Trans.
  Ultrason., Ferroelectr., Freq. Control}, vol.~39, pp. 262--267, 1992.

\bibitem{Jensen1996}
J.~Jensen, ``Field: A program for simulating ultrasound systems,'' in
  \emph{Medical \& Biological Engineering \& Computing}, vol.~34, no.~1, 1996,
  pp. 351--353.

\bibitem{Marmottant2005}
P.~Marmottant, S.~van~der Meer, M.~Emmer, M.~Versluis, N.~de~Jong,
  S.~Hilgenfeldt, and D.~Lohse, ``A model for large amplitude oscillations of
  coated bubbles accounting for buckling and rupture,'' \emph{The Journal of
  the Acoustical Society of America}, vol. 118, no.~6, pp. 3499--3505, 2005.

\bibitem{Christensen-Jeffries2017}
K.~Christensen-Jeffries, S.~Harput, J.~Brown, P.~N.~T. Wells, P.~Aljabar,
  C.~Dunsby, M.-X. Tang, and R.~J. Eckersley, ``Microbubble axial localization
  errors in ultrasonic super-resolution imaging,'' \emph{IEEE Trans. Ultrason.,
  Ferroelectr., Freq. Control}, vol.~64, no.~11, pp. 1644--1654, 2017.

\end{thebibliography}
\bibliographystyle{IEEEtran}

\end{document}